\documentclass[11pt,twoside]{article}
\usepackage{asp2010}

\resetcounters

\begin{document}

\title{GammaLib - A new framework for the analysis of Astronomical Gamma-Ray Data}
\author{J\"urgen Kn\"odlseder$^{1,2}$
\affil{$^1$Universit\'e de Toulouse; UPS-OMP; IRAP;  Toulouse, France}
\affil{$^2$CNRS; IRAP; 9 Av. colonel Roche, BP 44346, F-31028 Toulouse cedex 4, France}}

\begin{abstract}
With the advent of a new generation of telescopes (INTEGRAL, Fermi, H.E.S.S., MAGIC, 
VERITAS, MILAGRO) and the prospects of planned observatories such as CTA or HAWC, 
gamma-ray astronomy is becoming an integral part of modern astrophysical research. 
Analysing gamma-ray data is still a major challenge, and today relies on a large diversity
of tools and software frameworks that were specifically developed for each instrument.
With the goal of facilitating and unifying the analysis of gamma-ray data, we are currently
developing an innovative data analysis toolbox, called the GammaLib, that enables
gamma-ray data analysis in an instrument independent way.
We will present the basic ideas that are behind the GammaLib, and describe its
architecture and usage.
\end{abstract}

\section{Introduction}

The last decade has seen important progress in the field of gamma-ray astronomy, thanks
to significant improvements in the performance of ground-based and space-based gamma-ray 
telescopes \citep{vandenbroucke2010}.
Gamma-ray photons are nowadays studied over more than 8 decades in energy,
from a few 100 keV up to more than 10 TeV.
The technologies used for observing gamma-rays are very diverse, and cover
indirect imaging devices, such as coded mask or Compton telescopes,
and direct imaging devices, such as pair creation telescopes, and
air or water Cherenkov telescopes.

Despite this technical diversity, the high-level data produced by the instruments
for scientific analysis have great similarities.
Generally, the data are comprised of individual events that are characterised by an
arrival time, a direction (either given in celestial coordinates for direct imaging devices or
in detector coordinates for indirect imaging devices), and an energy estimate.
Standards, such as 
the FITS data format \citep{pence2010}
and the OGIP conventions \citep{corcoran1995},
have been implemented for many space-based instruments (CGRO, INTEGRAL, Fermi),
although ground-based telescopes (H.E.S.S., MAGIC, VERITAS, MILAGRO) still use their own 
collaboration internal data formats.
An increasing number of projects use analysis tools inspired by HEASARC's 
FTOOLS \citep{pence1993},
and implement the IRAF parameter interface \citep{valdes1992},
allowing for command line control and scripting.
Most (if not all) analysis methods imply at some point the fitting of parametric models,
employing maximum likelihood techniques to infer phyiscal parameters and their
uncertainties \citep{cash1979}.

Despite these common features, there exist no common tools yet for the scientific analysis of 
gamma-ray data.
So far, each instrument comes with its proper suite of software tools, which often requires costly
development cycles and maintenance efforts, and which puts the burden on the astronomer to 
learn how to use each of them for his studies.
For X-ray astronomy, HEASARC has developed standards (such as XSELECT or XSPEC)
that unify the data analysis tasks, making X-ray data more accessible to the astronomical
community at large.
We propose here to follow a similar approach for gamma-ray astronomy.

\section{What is GammaLib?}

GammaLib has the ambition to provide a unified framework for the high-level analysis of 
astronomical gamma-ray  data.
GammaLib, which we currently develop at IRAP (Toulouse, France), is a self-contained, 
instrument independent, open source, multi-platform C++ library that implements all code 
required for high-level science analysis of astronomical gamma-ray data.

Self-contained means that GammaLib does not rely on any third-party software, with the only
exception of the cfitsio library from HEASARC that is used to implement the FITS interface.
This makes GammaLib basically independent of any other software package,
increasing the maintainability and enhancing the portability of the library.
The drawback of this approach is that a lot of code had to be we written in a first place to
implement basic functionnalities in GammaLib, such as model fitting, IRAF and XML parameter
interfaces, or supporting WCS and HEALPix coordinate projections.

Instrument independent means that GammaLib potentially supports any gamma-ray astronomy
instrument.
Large parts of the code treat gamma-ray observations in an abstract representation, and do
neither depend on the characteristics of the employed instrument, nor on the particular
formats in which data and instrument response functions are delivered.
Instrument specific aspects are implemented as isolated and well defined modules that
interact with the rest of the library through a common interface.
Adding a new instrument to GammaLib consists of implementing a new instrument specific 
module, which considerably reduces the development costs compared to the full-fledged
development of an instrument-specific scientific analysis package.
This philosophy also enables the joint analysis of data from different instruments, providing
a framework that allows for consistent broad-band spectral fitting or imaging.

Open source means that the GammaLib source code is freely available under the GNU General
Public license version 3.
The latest source code can be downloaded from 
\url{https://sourceforge.net/projects/gammalib/}
which also provides bug trackers and mailing lists.
Further information and documentation on GammaLib can be found on
\url{http://gammalib.sourceforge.net/}.

Multi-platform means that GammaLib is designed to compile on any POSIX compliant platform.
So far, GammaLib has been successfully compiled and tested on Mac OS X, OpenBSD,
OpenSolaris (using the gcc compiler) and many Linux flavours.
We are currently setting up an automated continuous build system based on virtual machine
technology to guarantee multi-platform compliance throughout the continuing development 
process.
Pre-packed binary versions of the code will become available soon.

C++ library means that GammaLib makes heavily use of C++ classes.
Instrument independency is achieved by using abstract virtual base classes, which are
implemented as derived classes in the instrument specific modules.

\section{How is GammaLib organised?}

\begin{figure*}[!t]
\centering
\includegraphics[width=3.9in]{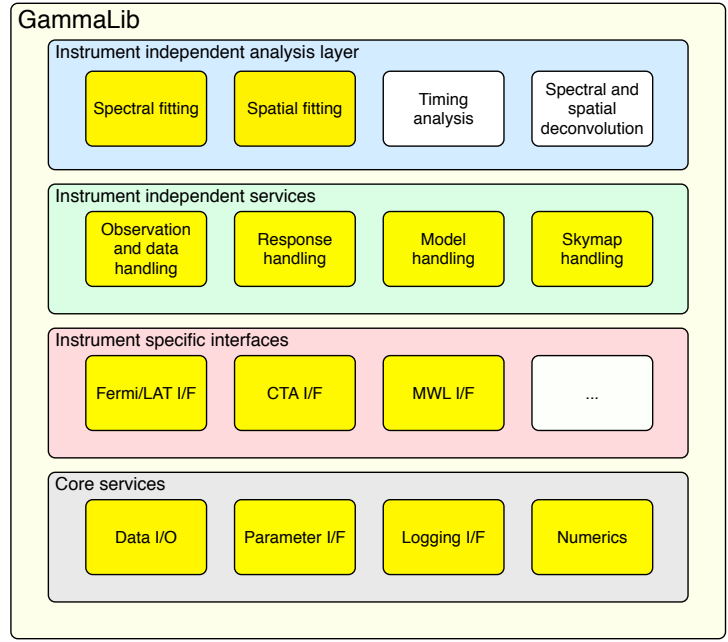}
\caption{
Organisation of GammaLib into software layers and modules.
Existing modules are shown in yellow, while planned modules are shown in white.
\label{fig:structure}
}
\end{figure*}

GammaLib is organised into four software layers, each of which comprises a number of modules
(see Figure \ref{fig:structure}).
The top layer provides support for instrument independent high-level data analysis, enabling 
spectral and spatial fitting by forwards folding of parametric source and background models.
Additional modules that are foreseen (but not yet implemented) will deal with timing analysis,
spectral and spatial deconvolutions.
The second layer provides an instrument independent interface to data products, comprising
information related to observations (event data, pointing and exposure information),
instrument response functions,
parametric source and background models used for model fitting, and
sky maps.
The latter may be defined using either WCS or HEALPix projections.
The modules of the third layer implement interfaces for specific gamma-ray instruments.
So far, the analysis of Fermi/LAT data is supported, and a prototype interface for CTA data has
been implemented.
Furthermore, a general multi-wavelength interface allows adding non-gamma-ray
flux points to spectral fits, enabling joint broad-band multi-wavelength
fitting of spectral energy distributions.
The forth layer provides core services for GammaLib.
They comprise
input and output of data (either in FITS format through the cfitsio library \citep{pence1999}
or in CSV format using a native interface),
handling of parameters (either in the IRAF parameter format or in XML format),
information logging (either to the console and/or into a dedicated log file),
and numerical algorithms and methods needed for data analysis
(e.g. optimization of functions, (sparse) matrix and vector handling, solving of linear equations, 
special mathematical functions, numerical integration and derivatives, interpolation of
data, random number generator).

\section{How can GammaLib be used?}

GammaLib implements an Application Programming Interface (API) that can be used
to bind functionalities into specific analysis executables.
In particular, GammaLib supports the creation of new FTOOLS, which are implemented
by deriving new classes from the {\tt GApplication} base class.
A very simple C++ executable that reads source model definitions from a file named
{\tt "source.xml"} and prints them to the console would look like this:
\begin{verbatim}
#include <iostream>
#include "GammaLib.hpp"
int main(void) {
    GModels models("source.xml");
    std::cout << models << std::endl;
    return 0;
}
\end{verbatim}

GammaLib C++ classes are also wrapped into Python using SWIG 
(see\break \url{http://www.swig.org/}),
providing a Python module that can be used to create analysis scripts or to perform
interactive analysis.
The same result as above in Python would be obtained by typing:
\begin{verbatim}
$ python
>>> from gammalib import *
>>> models = GModels("source.xml")
>>> print models
\end{verbatim}

We currently use GammaLib to implement a set of prototype FTOOLS\break
(\url{http://cta.irap.omp.eu/ctools/})
for the scientific analysis of data from the next generation ground-based Cherenkov
Telescope Array (CTA).
CTA is in the preparatory phase with an envisioned first light in 2015.


\bibliography{P069}

\end{document}